\documentclass{article}
\usepackage[accepted]{icml2018}
\usepackage{float}
\usepackage{latexsym}
\usepackage{amsfonts,amsthm,amssymb}
\usepackage{amsmath}
\usepackage{euscript}
\usepackage{amstext}
\usepackage{graphicx}
\usepackage{color}
\usepackage{nicefrac}
\usepackage{multirow}
\usepackage{lipsum}
\usepackage{enumitem}
\usepackage{microtype}
\usepackage{subfigure}
\usepackage{booktabs}
\usepackage{balance}

\newif\iffull
\fulltrue

\newif\ifshort
\shortfalse

\ifshort
\fi

\usepackage{amsmath}
\newtheorem{theorem}{Theorem}[section]

\newtheorem{lemma}[theorem]{Lemma}

\newtheorem{corollary}[theorem]{Corollary}

\newtheorem{definition}[theorem]{Definition}

\newcommand{\kcore}{k\text{-core}}
\def\polylog{\operatorname{polylog}}
\newcommand{\delete}[1]{}

\begin{document}
\twocolumn[
\icmltitle{Parallel and Streaming Algorithms for K-Core Decomposition}
\begin{icmlauthorlist}
	\icmlauthor{Hossein Esfandiari}{Go}
	\icmlauthor{Silvio Lattanzi}{Go}
	\icmlauthor{Vahab Mirrokni}{Go}
\end{icmlauthorlist}
\icmlaffiliation{Go}{Google Research}
\icmlcorrespondingauthor{Hossein Esfandiari}{esfandiari@googol.com}
\vskip 0.3in
]
\printAffiliationsAndNotice{}
\begin{abstract}
The $k$-core decomposition is a fundamental primitive in many  machine learning and data mining applications. We present the first  distributed and the first streaming algorithms to compute and maintain an approximate $k$-core decomposition with provable guarantees. Our algorithms achieve  rigorous bounds on space complexity while bounding the number of passes or number of rounds of computation. We do so by presenting a new powerful sketching technique for $k$-core decomposition, and then by showing it can be computed efficiently in both streaming and MapReduce models. Finally, we confirm the effectiveness of our sketching technique empirically on a number of publicly available graphs.
\end{abstract}

\section{Introduction}
A wide range of  data mining, machine learning and social network analysis problems can be modeled as graph mining tasks on large graphs. 
The ability to analyze layers of connectivity is useful to understand the hierarchical structure of the input data and  the role of nodes in different networks. A commonly used technique for this task is the $k$-core decomposition:  a $k$-core of a graph is a maximal subgraph where every node has induced degree at least $k$. $k$-core decomposition  has many real world applications from understanding dynamics in social networks~\cite{BKLRS12} to graph visualization~\cite{ADBV05}, from describing protein functions based on protein-protein networks \cite{ASMKK06} to computing network centrality measures \cite{HMA06}. $k$-core is also widely used as a sub-routine for community detection algorithms~\cite{CGSV12,MPPT15} or for finding dense clusters in graphs ~\cite{LRJA10,MPPT15}. As a graph theoretic tool $k$-core decomposition has been used to solve the densest subgraph problem~\cite{LRJA10,DBLP:journals/pvldb/BahmaniKV12,DBLP:conf/www/EpastoLS15,esfandiari2015applications}.

$k$-core is often use a feature in machine learning systems with applications in network analysis, spam detection and biology. Furthermore, in comparison with other density-based measure as the densest subgraph, it has the advantage to assign a score to every node in the network. Finally, the $k$-core decomposition induce a hierarchical clustering on the entire network.
For many applications in machine learning and in data mining, it is important to be able to compute it efficiently on large graphs.

In the past decade, with increasing size of data sets available in various applications, the need for developing scalable algorithms has become more important. By definition, the process of computing $k$-core decomposition is sequential: in order to find the $k$-core, one can keep removing all nodes of degree less than $k$ from the remaining graph until there is no such a node. As a result, computing $k$-cores for big graphs in distributed systems is a challenging task. In fact, while $k$-core decomposition has been studied extensively in the literature and many efficient decentralized and streaming heuristics have been developed for this problem~\cite{MPM13,SGJ15}, nevertheless developing a distributed or a streaming algorithm with provable guarantees for $k$-core decomposition problem remains an unsolved problem. One difficulty in tackling the problem is that simple non-adaptive sampling techniques used for similar problems as densest subgraph~\cite{LRJA10,esfandiari2015applications,DBLP:journals/pvldb/BahmaniKV12, DBLP:conf/www/EpastoLS15,bhattacharya2015space} do not work here (See Related Work for details). In this paper, we tackle this problem and present the first parallel and streaming algorithm for this problem with provable approximation guarantee. We do so by defining an approximate notion of $k$-core, and providing an adaptive space-efficient sketching technique that can be used to compute an approximate $k$-core decomposition efficiently. Roughly speaking, a $1-\epsilon$-approximate $k$-core is an induced subgraph that includes the $k$-core, and such that  the induced degree of every node is at least $(1-\epsilon)k$. 

{\noindent \bf Our Contributions.}
As a foundation to all our results, we provide a powerful sketching technique to compute a $1-\epsilon$-approximate $k$-core for all $k$ simultaneously. Our sketch is adaptive in nature and it is based on a novel iterative edge sampling strategy. In particular, we design a sketch of size $\tilde{O}(n)$ that can be constructed in $O(\log n)$ rounds of sampling\footnote{In the paper, we use the notation $\tilde{O}(\cdot)$ to denote the fact that poly-logarithmic factors are ignored.}.

We then show the first application of our sketching technique in designing a parallel algorithm for computing the $k$-core decomposition. More precisely, we present a MapReduce-based algorithm to compute a $1-\epsilon$ approximate $k$-core decomposition of a graph in $O(\log n)$ rounds of computations, where the load of each machine is $\tilde{O}(n)$, for any $\epsilon \in (0,1]$. 

Moreover, we show that one can implement our sketch for $k$-core decomposition in a streaming setting in \emph{one pass} using only $\tilde{O}(n)$ space. 
In particular, we present a  one-pass streaming algorithm for $1-\epsilon$-approximate $k$-core decomposition of  graphs with $\tilde{O}(n)$ space. 
\comment{We also extend our algorithm to the turnstile setting where we have both insertion and deletion of edges in the stream.}
\delete{We also extend our results to turnstile setting where we have both insertion and deletion of edges in the stream. We show how to maintain our sketch in turnstile setting using $\tilde{O}(n)$ space. However, we do not bound the update time. We leave the problem of bounding the update time in the presence of both insertions and deletions to future work.}
 
Finally, we show experimentally the efficiency and accuracy of our sketching algorithm on few real world networks.  

{\noindent \bf Related Work.}
The $k$-core decomposition problem is related to the densest subgraph problem.
Streaming and turnstile algorithms for the densest subgrpah problem have been studied extensively in the past~\cite{LRJA10,esfandiari2015applications,DBLP:journals/pvldb/BahmaniKV12,
DBLP:conf/www/EpastoLS15,bhattacharya2015space}. While these problems are related, the theoretical results known for the densest subgraph problem are not directly applicable to the $k$-core decomposition problem.

There are two types of algorithms for the densest subgraph problem in the streaming\delete{ and turnstile settings}. First type of algorithms simulates the process of iteratively removing vertices with small degrees~\cite{DBLP:journals/pvldb/BahmaniKV12,DBLP:conf/www/EpastoLS15,bhattacharya2015space}. All of these results are based on the fact that we only need logarithmic rounds of probing to find a $1/2$ approximation of the densest subgraph~\cite{DBLP:journals/pvldb/BahmaniKV12}. However, this can not be used to $1-\epsilon$ approximate the $k$-coreness numbers.

The second type of algorithms do a (non-adaptive) single pass and use uniform samplings of edges~\cite{esfandiari2015applications,MPPT15,mcgregor2015densest}. These results are based on the fact that the density of the optimum solution is proportional to the sampling rate with high probability, where the probability of failure  is \emph{exponentially} small. There are two obstacles toward applying this approach to approximating a $k$-core decomposition. First, by using uniform sampling it is not possible to obtain a $(1-\epsilon)$ approximation of the coreness number for nodes of constant degree (unless we do not sample all edges with probability one). Second, in order to achieve $\tilde{O}(n)$ space, we can only sample $\tilde{O}(1)$ edges per vertex. Hence, the probability that the degree of a vertex in the sampled is not proportional to the sampling rate, is not exponentially small anymore. Therefore it is not possible to union bound over exponentially many feasible solutions. To overcome this issue, we analyze the combinatorial correlation between feasible solutions and wisely pick polynomially many feasible solutions that approximate all of the feasible solutions. To the best of our knowledge this is the first work that analyzes the combinatorial correlation of different feasible solution on a graph.

In recent years the $k$-core decomposition problem received a lot of attention~\cite{BKLRS12,MPM13,ACCKU14,SGJ15, DBLP:conf/icde/ZhangYZQ17}, nevertheless we do not know of any previous distributed algorithms with small bounded memory and number of rounds.
A  recent related paper is ~\cite{SGJ15} where the authors present a streaming algorithm for the $k$-core decomposition problem. While the authors report good empirical results for their algorithm, they do not provide a guarantee for this problem, e.g., they do not prove an upper bound on the memory complexity of this algorithm. Finally we note that Monteresor et al.~\cite{MPM13} provide a distributed algorithm for this problem in a vertex-centric model. Although their model is different from our, more classic, MapReduce setting and their bound on the number of rounds is linear instead we achieve a logarithmic bound.

\section{Preliminaries}
\label{sec:prelim}

In this section, we introduce the main definitions and the computational models that we consider in the paper. We start
by defining $k$-core and by introducing the concept of approximate $k$-core. Then we describe the MapReduce and
streaming models.

\noindent{\bf Approximate $k$-core.} Let $G=(V, E)$ be a graph with $|V|=n$ nodes and $|E|=m$ edges. Let $H$ be a subgraph of $G$, for any node
$v\in G$ we denote by $d(v)$ the degree of the node in $G$ and for any node $v\in H$ we denote by $d_H(v)$
the degree of $v$ in the subgraph induced by $H$.  A $\kcore$
is a maximal subgraph $H\subseteq G$ such that $\forall v\in H$ we have $d_H(v)\geq k$. Note that for any $k$ 
the $\kcore$ is unique and it may be possibly disconnected. We say that a vertex $v$ has \emph{coreness} number $k$
if it belongs to the $k$-core but it does not belong to the $(k+1)$-core. We denote the coreness number of node $i$
in the graph $G$ with $C_G(i)$(we drop the subscript notation when the graph is clear from the context).

We define the \emph{core labeling} for a graph $G$
as the labeling where every vertex $v$ is labeled with its \emph{coreness} number. It is wroth noting that this
labeling is unique and that it defines a hierarchical decomposition of $G$.

\begin{figure}[h]
	\vskip -0.1in
\begin{center}
\includegraphics[width=1.9in]{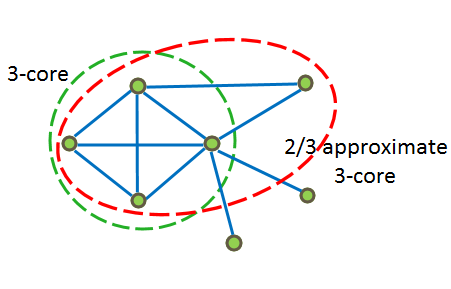}
	\vskip -0.1in
\caption{\label{fig:def}\small Example of $3$-core and $\frac23$-approximate 3-core.}
\end{center}
	\vskip -0.05in
\end{figure}

In this paper we are interested in computing a good approximation of the \emph{core labeling} for a graph 
$G$ efficiently in the MapReduce and in the streaming model. For this reason, we introduce the concept of
$1-\epsilon$ approximate $k$-core. We define a $1-\epsilon$ approximation to the 
$k$-core of $G$ to be a subgraph $H$ of $G$ that contains the $k$-core of $G$ and such that $\forall v\in H$
we have $d_H(v)\geq (1-\epsilon)k$. In other words, a $1-\epsilon$ approximation to the $k$-core of $G$ is 
a subgraph of the $(1-\epsilon)k$-core of $G$ and  supergraph of the $k$-core of $G$. In Figure~\ref{fig:def}
we present the 3-core for a small graph and a $\frac23$-approximate 3-core.

Similarly, a $1-\epsilon$ approximate core-labeling of a graph $G$ is a labeling of the vertices in $G$, where 
each vertex is labelled with a number between its coreness  number and its coreness number multiplied by
$\frac 1 {1-\epsilon}$.

In the paper  we often refer to the classic greedy algorithm~\cite{DBLP:journals/jacm/MatulaB83}(also known as peeling algorithm) to compute the coreness number. The algorithm works as follows: nodes are removed from the graph iteratively. In particular, in iteration $i$ of the algorithm all nodes with degree smaller or equal to $i$ are removed iteratively and they are assigned coreness number $i$. It is possible to show that the algorithm computes the correct coreness number of all nodes in the graph and it can be implemented in linear time.

\noindent{\bf MapReduce model.} Here we briefly recall the main aspect of the model by 
Karloff et al.~\cite{DBLP:conf/soda/KarloffSV10} of the MapReduce 
framework~\cite{DBLP:journals/cacm/DeanG10}.

In the MapReduce model, the computation happens in parallel in several rounds. In each round,
data is analyzed on each machine in parallel and then the output of the computations are shuffled 
between machines. The model has two main restrictions, one on the total number of machines and 
another on the memory available on each machine. More specifically, given an input of size $N$, 
and a small constant $\epsilon > 0$, in the model there are $N^{1-\epsilon}$ machines,
each with $N^{1 - \epsilon}$ memory available. Note that, the total
amount of memory available to the entire system is $O(N^{2-2\epsilon})$.

The efficiency of an algorithm is measured by the number of the ``rounds'' needed
by the algorithm to terminate. Classes of algorithms of particular interest are the ones that 
run in a constant or poly-logarithmic number of rounds.

\noindent{\bf Streaming \delete{and turnstile model.}} We also analyze the approximate core labelling problem
in the streaming model~\cite{DBLP:journals/tcs/MunroP80}. In this model the input consists
of an undirected graph $G=\left( V,E\right)$ and the input is presented as a stream of edges. 
The goal of our algorithm is to obtain a good approximation of 
the core labelling at the end of the stream using only small memory ($\tilde{O}(n)$). 

\comment{Finally, we also consider the turnstile model~\cite{DBLP:conf/soda/Muthukrishnan03} where the input consists
of a sequence of edges additions and deletions and 
the goal is again to design an algorithm to compute an approximation of 
the core labelling at the end of the stream using only small memory.}

\section{Sketching $k$-Cores}\label{sec:sketch}

In this section we present a sketch to compute an approximate core labelling that uses only $O(n \polylog(n))$ space. Compared with previous sketching for similar problems~\cite{LRJA10,esfandiari2015applications,DBLP:journals/pvldb/BahmaniKV12,
DBLP:conf/www/EpastoLS15,bhattacharya2015space} our sketching samples different area of the graphs with different, carefully selected, probabilities.

The main idea behind the sketch is to sample edges more aggressively in denser areas of the graph and less aggressively in sparser areas. 
More specifically, the algorithm works as follows: we start by sampling edges with some small probability, $p$, so that the resulting sampled graph, $H$, is sparse. We then compute the coreness numbers for the vertices in $H$. The key observation is that if a vertex has logarithmic coreness number in $H$ we can precisely estimate its coreness number in the input graph $G$. Furthermore we can show that if a vertex has large enough coreness number in the input graph $G$ it will have at least logarithmic coreness number in $H$. So using this technique we can detect efficiently all nodes with sufficiently high coreness number.
To compute the coreness numbers of the rest of the node in the graph, we first remove from the graph the nodes for which we have a good estimation and then we iterate the same approach. In particular we double our sample probability $p$ and sample edges again. Interestingly, we can show that by sampling edges adaptively, we can iteratively estimate the coreness of all nodes in the graph by analyzing only sparse subgraphs. 

We are now ready to describe our sketching algorithm in details. We start by describing a basic subroutine that estimates a modified version of the coreness number. We dubbed the subroutine \emph{ExclusiveCorenessLabeling}. The subroutine takes as input a subgraph, $H$, and a subset of the vertices $\Lambda \subseteq H$ and it runs a modified version of the classic peeling algorithm~\cite{DBLP:journals/jacm/MatulaB83} to compute the coreness number. The main difference between \emph{ExclusiveCorenessLabeling} and the peeling algorithm in~\cite{DBLP:journals/jacm/MatulaB83} is that we do not compute labels for nodes in $\Lambda$ and we do not remove them from the subgraph $H$. The pseudocode for \emph{ExclusiveCorenessLabeling} is presented in Algorithm~\ref{Alg:GreedyLambda}.
\begin{center}
\begin{algorithm}
\begin{algorithmic}[1]
\small
\STATE \textbf{Input:} A graph $H$ with $n$ vertices and a set $\Lambda\subseteq V_H$.
\STATE Initialize $\Gamma = V_H \setminus \Lambda$
\STATE Initialize $l\leftarrow 0$ 
\WHILE{$\Gamma \neq \emptyset$}
	\WHILE{$\min_{v\in \Gamma}(d_H(v))\leq l$}
		\STATE Let $v \leftarrow argmin_{v\in \Gamma}(d_H(v))$
		\STATE Set $l_v \leftarrow l$
		\STATE Remove $v$ from $\Gamma$
		\STATE Remove $v$ from $H$
	\ENDWHILE
	$l\leftarrow l+1$
\ENDWHILE
\caption{\small $ExclusiveCorenessLabeling(H,\Lambda)$}
\label{Alg:GreedyLambda}
\end{algorithmic}
\end{algorithm}
\end{center}

During the execution of the algorithm we use  subroutine \emph{ExclusiveCorenessLabeling} to compute a
labelling for the subset of the nodes in $H$ for which we do not have already a good estimate of the coreness number.

Now we can formally present our algorithm, we start by sampling the graph $G$  with $p\in O\left(\frac{\log n}{\epsilon^2 n}\right)$. In this way, we obtain a sparse graph $H_0$. Then we run \emph{ExclusiveCorenessLabeling} with $H=H_0$ and $\Lambda = \emptyset$ to
obtain a labeling of the nodes in $H_0$. Let $l_0(i)$ be the label of vertex $i$ in this labeling. If a vertex $i$ has $l_0(i)\geq C\log n$, for a specific constant $C>0$, we can estimate its coreness number in $G$ precisely. Intuitively this is true because we are sampling the edges independently so we can use concentration results to bound its coreness number. Hence, in the first round of our algorithm we can compute a precise estimate of the coreness number for all nodes $i$ with $l_0(i)\geq C\log n$.

In the rest of the execution of our algorithm we can recurse on the remaining nodes. To do so, we add the nodes with a good estimate to the set $\Lambda$ and we remove from $G$ the edges in the subgraph induced by the nodes in $\Lambda$. Then we increase the sampling probability $p$ by $2$ and sample $G$ again. Similarly we obtain a new subgraph $H_1$ and we run \emph{ExclusiveCorenessLabeling} with $H=H_1$ and $\Lambda$ equal to the current $\Lambda$. So we obtain a labeling $l_1$ for the nodes in $H_1\setminus \Lambda$. and also in this case if a vertex $i$ has $l_1(i)\geq C\log n$, for a specific constant $C>0$, we can estimate its coreness number in $G$ precisely.

We iterate this algorithm for $\log n$ steps. In the remaining of the section we first present pseudocode of our sketching algorithm(Algorithm~\ref{Alg:sketch}) then we show that at the end of the execution of the algorithm we have a good estimation of the coreness number for all nodes in $G$. Finally we argue that in every iteration the graphs $H_i$ are sparse so the algorithm uses only small memory at any point in time.

We start by providing the pseudocode of the algorithm in Algorithm~\ref{Alg:sketch}.


\begin{algorithm}
\begin{algorithmic}[1]
\small
\STATE \textbf{Input:} A graph $G$ with $n$ vertices and parameter $\epsilon \in (0,1]$.
\STATE Initialize $\Lambda \leftarrow \emptyset$
\STATE Initialize $p_0 \leftarrow \frac{96 \log n}{\epsilon^2 n}$
\FOR{$j = 0$ to $\log n$}
    \STATE Let $H_j$ be a subgraph of $G$ with the edges sampled independently with probability $p_j$
    \STATE Run $Exclusive\_Core\_Labeling(H_j,\Lambda)$ and denote the label of vertex $i$ on $H_j$ by $l_j(i)$
        \FOR{$i \in H_j$}
  	    \IF{$l_j(i) \geq \frac{192\log n}{\epsilon^2} \vee p_j = 1$}
		\STATE \textit{// Node $i$ has sufficiently high degree to estimate its coreness number.}
		\IF{$l_j(i) \leq \frac{384\log n}{\epsilon^2}$}
		    \STATE Set the label of vertex $i$ to $(1-\epsilon)\frac{l_j(i)}{p_j}$
		    \STATE Add $i$ to $\Lambda$
		\ELSE
		    \STATE Set the label of vertex $i$ to $ \frac{2(1-\epsilon)n}{2^{j-1}}$
		    \STATE Add $i$ to $\Lambda$
		\ENDIF
	    \ENDIF
	\ENDFOR
    \STATE Remove from $G$ the edges of $G$ induced by $\Lambda$
    \STATE $p_{j+1} \leftarrow 2 p_j$
\ENDFOR
\caption{\small A sketch based algorithm to compute $1-O(\epsilon)$ approximate core-labeling.}
\label{Alg:sketch}
\end{algorithmic}
\end{algorithm}

We are now ready to prove the main properties of our sketching technique. We start by stating two technical lemma whose proofs follow from application of concentration bounds and are presented in the appendix. The main goal of the lemma is to relate the degree of a vertex $v$ in a subgraph of $G$ and the sampled subgraph $H$.


\begin{lemma}\label{lm:cher0}
	Let $G$ be a graph and let $\epsilon \in (0,1]$ and $\delta\in (0,1)$ be two arbitrary numbers. Let $f(n)$ be a function of $n$ such that $f(n)\geq \frac{48 \log{\frac n{\delta}}}{\epsilon^2}$ and let $H$ be a  subgraph of $G$ that contains each edge of $G$ independently with probability $p\geq \frac{48 \log{\frac n{\delta}}}{\epsilon^2 f(n)}$. Then for all $v\in G$ the following statements holds, with probability $1-\frac {\delta}{3n^2}$:
		(i) If $d_G(v) \geq f(n)$ we have $|d_H(v) - p d_G(v)|\leq \epsilon d_H(v)$, 
		(ii) If $d_G(v) < f(n)$ we have $d_H(v) < 2 p f(n)$. 
\end{lemma}


\begin{lemma}\label{lm:cher2}
	Let $G$ be a graph and let $\epsilon \in (0,1]$ and $\delta\in (0,1)$ be two arbitrary numbers. Let $f(n)$ be a function of $n$ such that $f(n)\geq \frac{48 \log{\frac n{\delta}}}{\epsilon^2}$ and let $H$ be a  subgraph of $G$ that contains each edge of $G$ independently with probability $p\geq \frac{48 \log{\frac n{\delta}}}{\epsilon^2 f(n)}$. For all $v\in G$ the following statements holds, with probability $1-\frac {\delta}{3n^2}$:
		(i) If $d_H(v) \geq 2 p f(n)$ we have $|d_H(v) - p d_G(v)|\leq \epsilon d_H(v)$,
		(ii) If $d_H(v) < 2 p f(n)$ we have $d_G(v) \leq   2(1+\epsilon)f(n) $. 
	In addition, in the first case we have $d_G(v) \geq   2(1-\epsilon)f(n) $.
Furthermore, if the graph is directed the same claims hold for the in-degree($d^-(v)$) and the out-degree($d^+(v)$) of a node $v$.
\end{lemma}


In the remaining of this section we assume that Lemma~\ref{lm:cher0} and Lemma~\ref{lm:cher2} hold and using this assumption we prove the main properties of our algorithm.

We start by comparing the labels computed by Algorithm \ref{Alg:GreedyLambda} with the coreness number of its input graph. Recall that we denote the coreness number of node $i$ in the graph $G$ with $C_G(i)$. 

\begin{lemma}\label{lm:GoodGreed}
	Let $H=(V,E)$ be an arbitrary graph and let $\Lambda \subseteq V$ be an arbitrary set of vertices. Let $\hat{H}=(\Lambda, \hat{E})$ be an arbitrary graph on the set of vertices $\Lambda$, and let $H'= H \cup \hat{H}$. Let $l_v$ be the label computed by $ExclusiveCorenessLabeling(H,\Lambda)$. Then for each vertex $v\in V\setminus \Lambda$ we have:
	(i) $l_v \geq C_{H'}(v)$, 
	(ii) if $C_{H'}(v) \leq \min_{u\in\Lambda}\big(C_{H'}(u)\big)$, we have $l_v=C_{H'}(v)$.
\end{lemma}
\begin{proof}
	By definition of coreness number, if we iteratively remove all vertices with degree $C_{H'}(v)-1$  from $H'$, vertex $v$ is not removed from the graph. Furthermore note that Algorithm \ref{Alg:GreedyLambda} does not remove any vertex with degree more than $C_{H'}(v)-1$ unless $C_{H'}(v)-1 < l$. Thus, we have $l_v \geq C_{H'}(v)$ as desired.
	
	Note that, if we set $\Lambda = \emptyset$, Algorithm \ref{Alg:GreedyLambda} acts as the greedy algorithm that computes the coreness numbers.
	Moreover notice that if $C_{H'}(v) \leq \min_{u\in\Lambda}\big(C_{H'}(u)\big)$, the classic peeling algorithm does not removed any of the vertices in $\Lambda$ until it does not consider nodes with degree smaller or equal than $l$. Therefore we have $l_v \geq C_{H'}(v)$ which proves the second statement of the theorem.
\end{proof}



We are now ready to state the two main Lemma proving the quality of the solution computed by our sketching technique.

\begin{lemma}\label{lm:levelj}
For all $0\leq j\leq \log n$ such that $p_j \leq 1$ and for any node $v$ added to $\Lambda$ in round $j$ we have with probability $1-\frac{1}{3n}$ that: 
$C(v) < 2 (1+\epsilon)\frac{n}{2^{j-1}}$.

Furthermore for all $0\leq j\leq \log n$ such that $p_j < 1$ we have with probability $1-\frac{1}{3n}$ that: 
$C(v)\geq 2 (1-\epsilon)\frac{n}{2^{j}}  $.
\end{lemma}

\begin{lemma}\label{lm:acc}
	Algorithm \ref{Alg:sketch} computes a $1-2\epsilon$ approximate core labeling, with probability $1-\frac{2}{3n}$.
\end{lemma}

The proofs of the Lemma is presented in the appendix.


Now we give a lemma that bounds the total number of edges used in sketches $H_0,H_2,\dots,H_{\rho}$.

\begin{lemma}\label{lm:space}
	The number of edges in $\cup_{i=0}^{\rho} H_i$ produced by Algorithm \ref{Alg:sketch} is upper bounded by $O\left(\frac{n\log^2 n}{\epsilon^2}\right)$, with probability $1-\frac{1}{n}$.
\end{lemma}
\begin{proof}
In the proof, we assume that  the statement of Lemma \ref{lm:levelj} holds, and the statements of Lemma \ref{lm:cher0} and \ref{lm:cher2} hold for $H_{j,k}$ and $H_{j,v}$ for all choices of $j$ and $k$ and $v$.

Consider an arbitrary $0\leq j\leq \log n$. From Lemma \ref{lm:levelj}, we have that for any $v\in H_j\setminus (\cup_{i=0}^{j-1}\Lambda_j$) the coreness number of $v$ is bounded by $2(1+\epsilon)\frac{n}{2^{j-1}}$. Now consider an orientation of the edges of $H_j$ where every edge is oriented to its endpoint of smallest core number, breaking the ties in such a way that the in-degree of every node $v$, $d^-(v)$ is upperbounded by $C(v)$\footnote{Note that such an orientation exists, in fact  it can be obtained by orienting every edges to its endpoint that is first removed by the classic peeling algorithm used to compute the coreness number.}. Furthermore note that every edge in $H_j$ is incident to a node of coreness number at most $2(1+\epsilon)\frac{n}{2^{j-1}}$, so using Lemma \ref{lm:cher0} we have that in-degree of every node in $H_j$ is bounded by $2(1+\epsilon)^2\frac{n}{2^{j-1}}p^j= 384\frac{(1+\epsilon)^2}{\epsilon^2}\log n$. So summing over all the in-degrees we get that the  number of edges in $H_j$ is bounded by $384 \frac{(1+\epsilon)^2}{\epsilon^2} n \log n$. We conclude the proof by noticing that there are at most  $\log n$ different $H_j$ so the total memory used is $384 \frac{(1+\epsilon)^2}{\epsilon^2} n \log^2 n$.
\end{proof}

Putting together Lemma~\ref{lm:acc} and Lemma~\ref{lm:space} we get the main theorem of this section.
\begin{theorem}\label{th:sketch}
	Algorithm \ref{Alg:sketch} computes a $1-2\epsilon$ approximate core labeling and the total spaced used by the algorithm is $O\left(\frac{n\log^2 n}{\epsilon^2}\right)$, with probability $1-\frac{2}{n}$.
\end{theorem}

\section{MapReduce and Streaming Algorithms}
In this section we show how to compute our sketch efficiently using a MapReduce or a streaming algorithm.

\subsection{MapReduce algorithm}

Here, we show how to use implement the sketch introduced in Section~\ref{sec:sketch} in the MapReduce model. In
this way we obtain an efficient MapReduce algorithm for dense graphs\footnote{It is important to note that we only use
polylogarithmic memory for each machine so our algorithm works also in more restrictive parallel models as the massively parallel
model~\cite{DBLP:conf/stoc/AndoniNOY14,DBLP:conf/stoc/ImMS17}}.  

Recall that the main limitation of the MapReduce model is on the number of machines and on the 
available memory on each machine.
Our algorithm runs for $2\log n$ rounds.\footnote{The algorithm can be implemented using 
$\log n$ MapReduce rounds, but for simplicity, here we present a $2\log n$ rounds version.}
In the first round of MapReduce, the edges are sampled in parallel 
with probability $p_0 = \frac{96 \log n}{\epsilon^2 n}$. In this way, we obtain a graph $H_0$ that we analyze
in the second round in a single machines(note that we can do it because from Lemma~\ref{lm:space}
we know that for all $i$ the number of edges in $H_i$ is bounded by $O\left(\frac{n\log^2 n}{\epsilon^2}\right)$. At the
end of the second round, we obtain the labeling for the nodes with high coreness number and we add them to  the set $\Lambda_0$. In the third round
we send the set $\Lambda_0$ to all the machined and we sample in parallel the edges in $|E|\setminus \Lambda_0$
with probability $2p_0$ in a round of MapReduce. In this way, we obtain $H_1$ that in the fourth round is analyzed by a single machine to obtain the labelling of few additional nodes that are added to $\Lambda_1$. By iterating this process for
$2\log n$ rounds, we obtain an approximation of the coreness number for each node. The pseudo-code for the MapReduce
algorithm is presented in Algorithm~\ref{Alg:mr}. 

\begin{algorithm}[t!]
\begin{algorithmic}[1]
\small
\STATE \textbf{Input:} A graph $G$ with $n$ vertices and parameter $\epsilon \in (0,1]$.
\STATE Initialize $\Lambda \leftarrow \emptyset$
\STATE Initialize $p_0 \leftarrow \frac{96 \log n}{\epsilon^2 n}$
\FOR{$j = 0$ to $\log n$}
    \STATE \textit{// First round of MapReduce}
    \STATE Send $\Lambda$ to all machines
    \STATE Let $E'$ be the set of edges of $G$ that are not contained in the graph induced by $\Lambda$ on $G$
    \STATE Sample with probability $p_j$ in parallel using $n$ machines the edges in $E'$
    \STATE \textit{// Second round of MapReduce}
    \STATE Send all the sampled edge to a single machine
    \STATE Let $H_j$ be the sampled subgraph of $G$
    \STATE Run $Exclusive\_Core\_Labeling(H_j,\Lambda)$ and denote the label of vertex $i$ on $H_j$ by $l_j(i)$
        \FOR{$i \in H_j$}
  	    \IF{$l_j(i) \geq \frac{192\log n}{\epsilon^2} \vee p_j = 1$}
		\STATE // Node $i$ has sufficiently high degree to estimate its coreness number.
		\IF{$l_j(i) \leq \frac{384\log n}{\epsilon^2}$}
		    \STATE Set the label of vertex $i$ to $(1-\epsilon)\frac{l_j(i)}{p_j}$
		    \STATE Add $i$ to $\Lambda$
		\ELSE
		    \STATE Set the label of vertex $i$ to $ \frac{2(1-\epsilon)n}{2^{j-1}}$
		    \STATE Add $i$ to $\Lambda$
		\ENDIF
	    \ENDIF
	\ENDFOR
    \STATE $p_{j+1} \leftarrow 2 p_j$
\ENDFOR
\caption{\small A MapReduce algorithm to compute $1-O(\epsilon)$-approximate core-labeling.}
\label{Alg:mr}
\end{algorithmic}
\end{algorithm}

By Theorem~\ref{th:sketch} presented in the previous section
 we obtain the following corollary.

\begin{corollary}\label{cor:mr}
	Let $G=(V,E)$ be a graph such that $|E|\in\Omega(|V|^{1+\gamma})$, for some constant $\gamma>0$. Then
	there is an algorithm that  computes w.h.p. an approximate core-labeling of the graph in the MapReduce model using
	$O(\log n)$ rounds of MapReduce.
\end{corollary}

\subsection{Semi-streaming algorithms}

Next we show an application of our sketch in the streaming setting. We  consider the setting where edges are only added to the graph. The main idea behind our streaming algorithm is to maintain at any point in time the sketch presented in Section~\ref{sec:sketch}, which requires only $\tilde{O}(n)$ space. In the remaining of the section we describe how we can maintain the sketching in streaming.

When an edge is added to $G$, we check by sampling if it is $H_0$. In this case in $H_0$, we recompute the labeling of $H_0$ and if one of the endpoints of the edge is added to $\Lambda_0$, we update the rest of the sketch to reflect this change. Then, if both endpoints of the edge are not contained in $\Lambda_0$, we check by sampling if the edge is contained in $H_1$. Also in this case, if it is in $H_1$, we recompute the labeling of $H_1$ and modify the sketch accordingly. We continue this procedure until both endpoints of the edge are contained in $\Lambda$. Notice that, by inserting edges $\cup_{i\leq j} \Lambda_i$s may only grow. Hence if at some point both endpoints of an edge $(u,v)$ are in $\cup_{i\leq j} \Lambda_i$, by inserting more edge $u$ and $v$ remain in $\cup_{i\leq j} \Lambda_i$.

The pseudo-code for the streaming 
algorithm is presented in Algorithm~\ref{Alg:streaming}(Note that here for simplicity we recompute the core labels after the insertion of an edge $(u,v)$. However, one might recurse over the neighborhood of $u$ and $v$ and update the core labels locally). 

By Theorem~\ref{th:sketch} presented in the previous section
 we obtain the following corollary.

\begin{corollary}\label{cor:streaming}
	There exists a streaming algorithm that computes w.h.p. an $(1-\epsilon)$-approximate core-labeling of the input graph using $O\left(\frac{n\log^2 n}{\epsilon^2}\right)$ space.
\end{corollary}

\begin{algorithm}[t!]
\begin{algorithmic}[1]
	\small
	\STATE \textbf{Input:} Stream of edges and parameter $\epsilon \in (0,1]$.
	\STATE Initialize $\Lambda_{i} \leftarrow \emptyset$, $\forall i$
         \STATE Initialize $p_0 \leftarrow \frac{96 \log n}{\epsilon^2 n}$
	 \STATE \textbf{Insertion of $(u,v)$}
	\STATE $r \leftarrow $ random number from $[0,1]$
	\FOR{$j = 0$ to $\log n$}
		\IF{$v\notin\cup_{i<j}\Lambda_{i}$ or $u \notin\cup_{i<j} \Lambda_{i}$}
			\IF{$r\leq p_j$}
				\STATE Add $(u,v)$ to $H_j$
				\STATE Run $Exclusive\_Core\_Labeling(H_j,\Lambda_{i})$ and denote the label of vertex $i$ on $H_j$ by $l_j(i)$
				\FOR{$i \in H_j$}
					\IF{$l_j(i) \geq \frac{192\log n}{\epsilon^2} \vee p_j = 1$}
						\IF{$l_j(i) \leq \frac{384\log n}{\epsilon^2}$}
							\STATE Set the label of vertex $i$ to $(1-\epsilon)\frac{l_j(i)}{p_j}$
							\STATE Add $i$ to $\Lambda_{j}$
						\ELSE
							\STATE Set the label of vertex $i$ to $ \frac{2(1-\epsilon)n}{2^{j-1}}$
							\STATE Add $i$ to $\Lambda_{j}$
						\ENDIF
					\ENDIF
				\ENDFOR
				\FOR{$j'=j+1$ to $\log n$}
					\STATE Remove from $H_{j'}$ any edge induced by $\Lambda_{j}$\;
				\ENDFOR
			\ENDIF
		\ELSE
			\STATE Break
		\ENDIF
		\STATE $p_{j+1}\leftarrow 2p_j$
	\ENDFOR
\caption{\small A streaming algorithm to compute $1-O(\epsilon)$ approximate core-labeling.}
\label{Alg:streaming}
\end{algorithmic}
\end{algorithm}

\section{Experiments}
In this section, we analyze the performances of our sketch in practice. First we describe our datasets. Next we discuss the implementations of our sketch presented in Section~\ref{sec:sketch} and our MapReduce algorithm. Then we study the scalability and the accuracy of our sketch. In particular, we analyze the trade-off between quality of the approximation and space used by the sketch.

\paragraph{Datasets.} We apply our sketch to eight real-world graphs available in the SNAP, Stanford Large Network Dataset Library~\cite{leskovec2016snap}: Enron~\cite{klimt2004introducing}, Epinions~\cite{DBLP:conf/semweb/RichardsonAD03}, Slashdot~\cite{DBLP:journals/im/LeskovecLDM09}, Twitter~\cite{DBLP:conf/nips/McAuleyL12}, Amazon~\cite{DBLP:journals/kais/YangL15}, Youtube~\cite{DBLP:journals/kais/YangL15}, LiveJournal~\cite{DBLP:journals/kais/YangL15}, Orkut~\cite{DBLP:journals/kais/YangL15} with respectively 36692, 75879, 82168, 81306, 334863, 1134890, 3997962 and 3072441  nodes and 183831, 508837, 948464, 1768149, 925872, 2987624, 34681189 and 117185083 edges.

\paragraph{Implementation details.} In order to have an efficient implementation of our sketch, we modify Algorithm~\ref{Alg:sketch} slightly. More specifically, we change line 14 to ``if $l_j(i) \geq T \vee p_j = 1$'' where $T$ is a parameter of our implementation. Furthermore, we also modify line 22 to ``$p_{j+1}\leftarrow M\cdot p_j$,  where $M$ is modifiable multiplicative factor (that in Algorithm~\ref{Alg:sketch} is fixed to 2).  
We also slightly modify our MapReduce algorithm to remove iteratively in parallel all nodes with degree smaller than $3$ before sending the remaining graph to a single machine.

\paragraph{Metrics.} 
To study the scalability of the algorithm we implement our MapReduce algorithm in distributed setting and we analyze the running time on different graphs by using a fixed number of machine.
To evaluate the quality of our sketch, we consider the quality of the approximation and the space used.

For the quality of the approximation, we report the median error and the error at the 60, 70, 80 and 90 percentile of our algorithm. In interest of space, we report the errors only on nodes with coreness number at least $5$, because high coreness number are harder to approximate and for almost all the nodes of coreness smaller than $5$ we have errors close to $0$.

For space we consider the maximum size of any sample graph $H_i$ and the sum of their sizes. Note that the first quantity bounds the memory used by our distributed algorithm or a multi-pass streaming algorithm, and the second one bounds the memory used by a single pass streaming algorithm.

\begin{figure}[ht!]
 		\vskip -0.1in
\begin{center}
\includegraphics[width=0.27\textwidth]{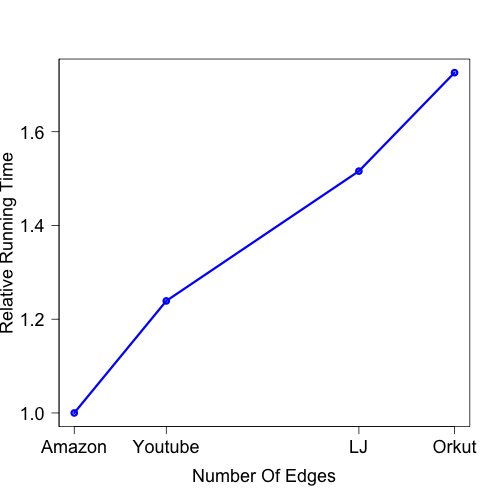}
	\vskip -0.05in
\caption{Running time of the distributed algorithm on graphs of increasing size.}
\label{fig:s}
\end{center}
	\vskip -0.05in
\end{figure}

 \paragraph{Scalability Results.} In Figure~\ref{fig:s} we present the results of our scalability experiments(In the experiment we fix $T=4$ and $M=2$). On the $x$ axis we order the graphs based on their number of edges, in the $y$ axis we show the relative running time on different graphs. Note that in the Figure the $x$ axis is in logscale and the $y$ axis is in linear scale so the running time of our algorithm grows sublinearly in the number of edges in the graph proving that our algorithm is able to leverage parallelization to obtain good performance.
 
For comparison we also run a simple iterative algorithm to estimate the k-core number that resembles a simple adaptation of the algorithm presented in~\cite{LRJA10,esfandiari2015applications,DBLP:journals/pvldb/BahmaniKV12, DBLP:conf/www/EpastoLS15,bhattacharya2015space} for densest subgraph. The adapted algorithm works as follows: it removes all nodes below a threshold $T$(initially equal to $4$) from the graphs in parallel and estimate their coreness number as $T$. Then when no node with degree smaller than $T$ is left, it iteratively increases $T$ by a multiplicative factor $M=2$ and recurse on the remaining graph. Interestingly we observe that this adapted algorithm is an order of magnitude slower than our distributed algorithm and so we could run it only on relatively small graphs like Amazon.\footnote{Note that the parallel version of the simple iterative algorithm is particularly slow in practice because it needs several parallel rounds to complete.
}

 \paragraph{Accuracy Results.} All the reported number are the average over $3$ runs of our algorithm. In all our experiment we either fix $T=3$ and vary $M$ or fix $M$ to $2$ and vary $T$. In Table~\ref{table:spaceT} we present the space used in our algorithm when we vary the value of $T$. 

 \begin{table}[h]
 \center
\scriptsize
  \caption{Number of edges stored by the sketch as a function of $T$}
    \begin{tabular}{|c|c|c|c|c|}
      \toprule
     Graph	&T=2 ($\max$)&T=3 ($\max$) &T=4 ($\max$)&T=5 ($\max$)\\
     \midrule
      Enron & 59300  & 93482  & 116110  & 142557 \\
      Epinions & 80791 & 120193  & 147513  & 169037  \\
      Slashdot & 132308  & 203789  & 258083  &  302763\\
      Twitter  &  58967 & 107164  &  148610 & 197321 \\
     \midrule
     &T=2 ($\sum$)&T=3 ($\sum$) &T=4 ($\sum$)&T=5 ($\sum$)\\
     \midrule
      Enron &  229549  &  337574 & 413380  & 470765  \\
      Epinions & 322622  &  436049 & 515461  &  575731\\
      Slashdot & 537006  & 799586  &  975408 & 1118110 \\
      Twitter  &  299734 & 501805  & 682251  & 848405 \\
          \bottomrule
  \end{tabular}
\label{table:spaceT}
\end{table}

There are few interesting things to note. First, the size of the maximum sampled graph is always significantly smaller than the size of input graph and in some cases it is more than one order of magnitude smaller (for example in the Twitter case). Interestingly, note that the relative size of the maximum sampled graph decrease with the size of the input graph. This suggests that the sketch would be even more effective when applied to a larger graph. 
Also  the total size of the sketch is also smaller than the size of the graph in many cases (for instance, the sketch for Twitter is always smaller than half of the size of the input graph). 
This implies that we can compute an approximation of the coreness number without processing most of the edges in the input graph.

\begin{figure}[ht!]
\begin{center}
		\vskip -0.1in
	
\subfigure[Enron]{\includegraphics[width=0.23\textwidth,keepaspectratio]{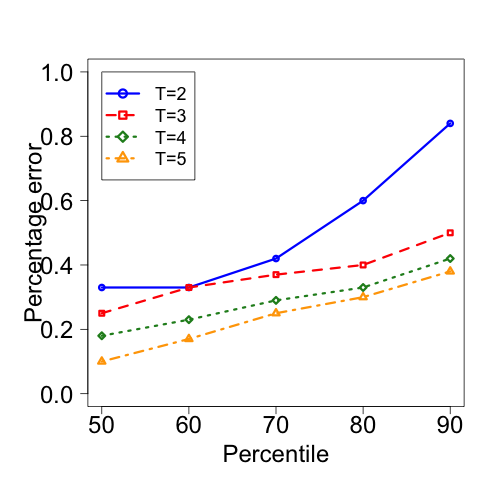}}
\subfigure[Epinions]{\includegraphics[width=0.23\textwidth,keepaspectratio]{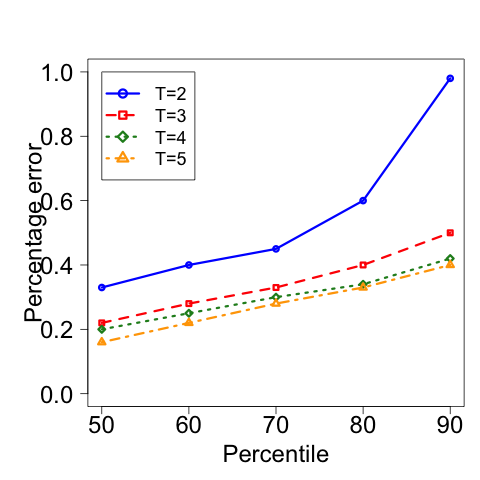}}\\

	\vskip -0.1in

\subfigure[Slashdot]{\includegraphics[width=0.23\textwidth,keepaspectratio]{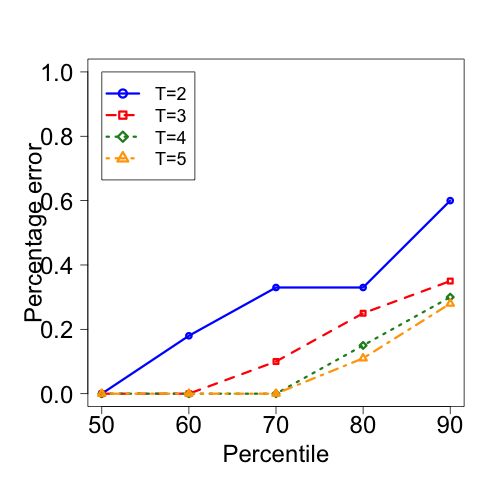}}
\subfigure[Twitter]{\includegraphics[width=0.23\textwidth,keepaspectratio]{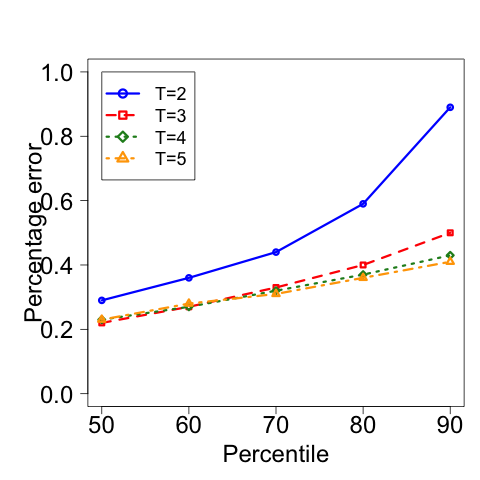}}
	\vskip -0.05in

\caption{The approximation error of our sketch. We show the error at median 60,70,80 and 90 percentile as a function of $T$ when we restrict our attention to node with coreness number at least $5$.}
\label{fig:th}
	\vskip -0.1in

\end{center}
\end{figure}

In 
Figure~\ref{fig:th}, we report the approximation error of our algorithm. First we note that as $T$ increases the approximation error decreases as predicted by our theorems. It is also interesting to note that the median error is always below $50\%$ and with $T\geq 3$ is below $25\%$. Observe for  $T\geq 3$, the error at the $90$ percentile is below $50\%$. Overall our sketch provides a good approximation of the coreness numbers.


Now we focus on the effect of $M$ on our sketch.  In Table~\ref{table:spaceM}, we present the space used by our algorithm as a function of $M$. Note that the maximum size of a single sample graph decrease with $M$, but the total size of the sketch increases (this is due to the increased number of sampled graphs). This suggests that we should use small $M$ in distributed settings where we have tighter space constraint  and larger $M$ when we want to design single pass streaming algorithms.

\begin{table}[h]
		\vskip -0.09in
\center
\scriptsize
  \caption{Number of edges stored by the sketch as a function of $M$}
    \begin{tabular}{|c|c|c|c|c|}
      \toprule
     Graph	&M=1.2 ($\max$)&M=1.4 ($\max$) &M=1.6 ($\max$)&M=2 ($\max$)\\
     \midrule
      Enron & 52202  & 67611  & 79075  & 85013 \\
      Epinions & 93059 & 101692  & 112514  & 122134  \\
      Slashdot & 128649  & 154429  & 171347  &  193257\\
      Twitter  &  51774 & 67314  &  81233 & 92842 \\
     \midrule
     &M=1.2 ($\sum$)&M=1.4 ($\sum$) &M=1.6 ($\sum$)&M=2 ($\sum$)\\
     \midrule
      Enron &  740240  &  485841 & 398151  & 355671  \\
      Epinions & 1023529  &  644102 & 517660  &  455226\\
      Slashdot & 1759205  & 1146974  &  938933 & 837956 \\
      Twitter  &  903449 & 650040  & 561669  & 521174 \\
          \bottomrule
  \end{tabular}
\label{table:spaceM}
\end{table}

\begin{figure}[ht!]
\begin{center}
		\vskip -0.1in
	
\subfigure[Enron]{\includegraphics[width=0.23\textwidth,keepaspectratio]{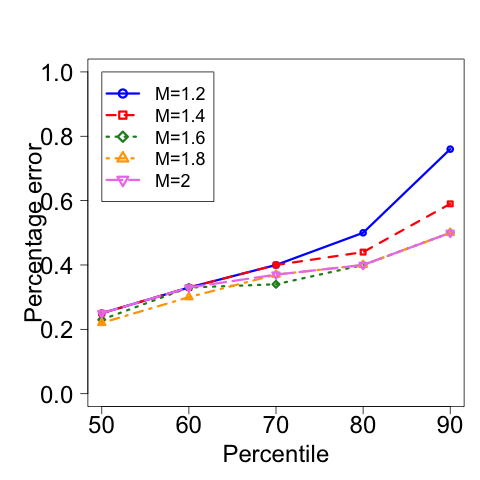}}
\subfigure[Epinions]{\includegraphics[width=0.23\textwidth,keepaspectratio]{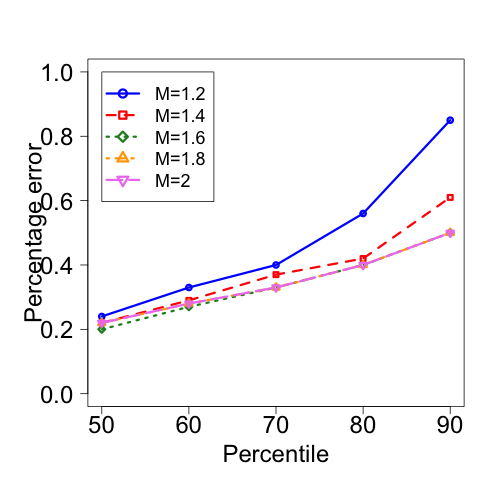}}\\

	\vskip -0.1in

\subfigure[Slashdot]{\includegraphics[width=0.23\textwidth,keepaspectratio]{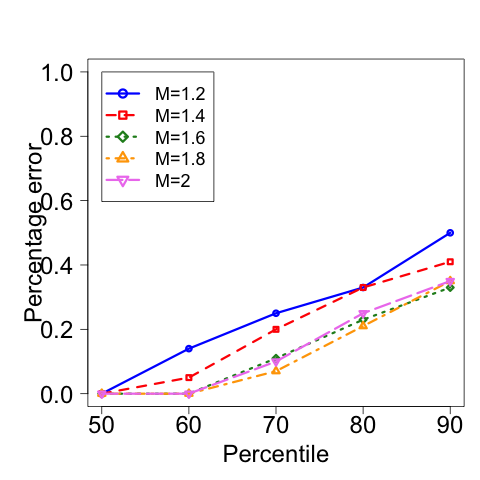}}
\subfigure[Twitter]{\includegraphics[width=0.23\textwidth,keepaspectratio]{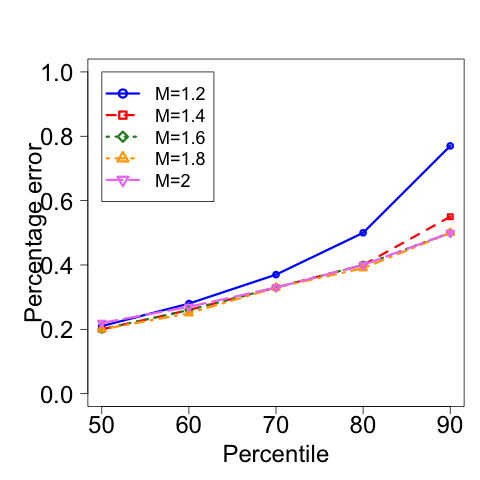}}
	\vskip -0.05in

\caption{The approximation error of our sketch. We show the error at median 60,70,80 and 90 percentile as a function of $M$ when we restrict our attention to node with coreness number at least $5$.}
	\vskip -0.1in

\label{fig:m}
\end{center}
\end{figure}

Finally it is interesting to note that as shown in Figure~\ref{fig:m}, the quality of the approximation is not very much influenced by the scaling factor $M$.

\section{Conclusions and future works}
In this paper we introduce a new sketching technique for computing the core-labeling of a graph. In particular, we design efficient MapReduce and streaming algorithms to approximate the coreness number of all the nodes in a graph efficiently. We also confirm the effectiveness of  our sketch via an empirical study.
The most interesting open problem in the area is to design a fully dynamic algorithm~\cite{italiano1999dynamic} to maintain the core-labeling of a graph by using only $\polylog n$ operations per update(edge addition or deletion).

\bibliographystyle{icml2018}
\balance
\bibliography{kcore}

\newpage
\section*{Appendix}
\begin{appendix} 
\section{Concentration bounds}\label{sec:app:conc}

Before giving a formal proof of Lemma~\ref{lm:cher0} and Lemma~\ref{lm:cher2} we recall a useful form of the Chernoff bound(for an exhaustive treatment on concentration of measure look at~\cite{DBLP:books/daglib/0025902}).

\begin{theorem}[Chernoff bound]
Let $X = \sum_{i=1}^n X_i$ where $X_i$, for $i\in [n]$ are independently distributed random variable in $[0,1]$. Then, for $0<\epsilon<1$, we have that
$$Pr[|X-E[X]|>\epsilon E[X]]\leq 2 exp\left(-\frac{\epsilon^2E[X]}{3}\right)$$
\end{theorem}

Now we are ready to prove our Lemma.

\begin{proof} (Proof of Lemma~\ref{lm:cher0})
	To prove this lemma we show that for a fixed vertex $v\in G$, the statement of the lemma holds with probability $1-\frac {\delta}{3n^3}$. Then by applying the union bound we obtain the lemma. 
	
	Let $X$ be the degree of a vertex $v$ in $H$. Note that each neighbor of $v$ in $G$ exists in $H$ with probability $p$. Thus we have $E[X]= p d_G(v)$. 
	
	First assume $d_G(v)\geq f(n)$. By the Chernoff bound we have
	\begin{align*}
	&Pr\Big( |X-E[X]| \geq \epsilon X \Big)  \\
	&\leq \footnotemark Pr\Big( |X-E[X]| \geq \frac{\epsilon}{2} E[X] \Big) \\
	&\leq 2 \exp(-\frac{ \epsilon^2 E[X]}{12})\\
	& = 2 \exp(-\frac{ \epsilon^2 p d_G(v)}{12}) \\
	& \leq 2 \exp(-\frac{ \epsilon^2 p f(n)}{12})\\
	& \leq 2 \exp(-4\log{\frac n{\delta}})\\
	&  = \frac {2\delta^4}{n^4} \leq \frac {\delta}{3n^3}
	\end{align*} 
	\footnotetext{ $X-E[X] \geq  \epsilon X$ implies $X-E[X] \geq \frac{\epsilon}{2} E[X]$ since $\frac {\epsilon} {1-\epsilon} \geq \frac{\epsilon}{2}$. Similarly, $E[X] - X \geq  \epsilon X$ implies $E[X] -X \geq \frac{\epsilon}{2} E[X]$ since $\frac {\epsilon} {1+\epsilon} \geq \frac{\epsilon}{2}$.}
	
	Assuming $n\geq 6$. This proves the first statement of the theorem.

	Next assume $d_G(v)< f(n)$. Again by the Chernoff bound we have
	\begin{align*}
	Pr\Big( X \geq 2 p f(n) \Big) &\leq 	Pr\Big( X-E[X] \geq  p f(n) \Big) \\
	&\leq  \exp\Big(-\frac{\big(pf(n)\big)^2}{3 E[X]}\Big) \\
	&\leq  \exp\Big(-\frac{pf(n)}{3}\Big) \\
	&\leq \exp\Big(-\frac{16\log{\frac n{\delta}}}{\epsilon^2}\Big) \\
	&\leq \frac {\delta}{3n^3}.
	\end{align*}
	This proves the second statement of the theorem and completes the proof for undirected graphs.
	
\end{proof}

\begin{proof} (Proof of Lemma~\ref{lm:cher2})
	To prove this lemma we show that whenever Lemma \ref{lm:cher0} holds, the statements of this lemma holds as well. Pick a vertex $v$. First assume $d_G(v)\geq f(n)$. In this case, using the first statement of Lemma \ref{lm:cher0} we have $|d_H(v) - p d_G(v)|\leq \epsilon d_H(v)$. Furthermore by Lemma \ref{lm:cher0} we know that if $d_H(v) \geq 2 p f(n)$  we have $d_G(v)\geq f(n)$. Those two facts together directly show the first statement of this lemma for vertex $v$. 

Moreover, suppose that $d_G(v)\geq f(n)$ then if $d_H(v) < 2 p f(n)$ we have 
	$
	p d_G(v)  \leq 	d_H(v) + \epsilon d_H(v) = (1+\epsilon)d_H(v)  <  (1+\epsilon) 2 p f(n).
	$
	Thus, we have $d_G(v) \leq 2(1+\epsilon)f(n) $, which shows the second statement of the lemma for vertex $v$.
	
	Otherwise, assume $d_G(v)< f(n)$. By Lemma \ref{lm:cher0} we know that  $d_H(v) < 2 p f(n)$. Thus, in this case the condition of the first statement of the lemma does not hold. Moreover, we have $d_G(v)< f(n) \leq  2(1+\epsilon)f(n)$, which shows the second statement of the lemma.
	
	Finally note that when $d_H(v) \geq 2 p f(n)$ we have
	$
	 p d_G(v) \geq d_H(v) -\epsilon d_H(v) \geq (1-\epsilon) 2 p f(n).
	$
	Thus, when $d_H(v) \geq 2 p f(n)$ we have $d_G(v)\geq 2(1-\epsilon) f(n)$.
\end{proof}

\section{An Example That Requires $\Omega(n)$ Rounds of Probing}\label{app:BadExample}
Here we show that $\Omega(n)$ rounds of probing is required to $1-\epsilon$ approximate coreness numbers in a graph.

Consider the following graph of $n$ vertices. For any $i$ between $0$ and $(n-5)/5$ and $j\in\{1,2,3\}$, vertex $5i+j$ is connected to vertex $5(i)+4$, $5(i+1)$, $5(i+1)+j$, furthermore for any $i$ between $0$ and $(n-1)/5$ the vertex $5i+4$ is connected with the vertex $5(i+1)$. Finally the  last $5$ vertices form  a $5$ clique, $K_5$.
Note that in this graph node $0$ has coreness number $0$, then all the nodes but the last $5$ have coreness number $3$ and the last $5$ have coreness number $4$.
Now we show that for any choice of $d$, if we iteratively remove vertices of degree less than $d$ for $o(n)$ rounds, it does not approximate the coreness number up to a $1-\epsilon$ factor. 

If $d<4$, it contains all node but vertex $0$ so it cannot be used to distinguish between nodes with coreness number $3$ or $4$. For $d=4$, it starts with the full graph and after round $i$ it is the graph induced on vertex with index bigger or equal than $5i$. Hence in order to distinguish nodes with coreness number $3$ or $4$, it requires $(n-5)/5$ rounds of probing. Finally, if $d\geq 5$ all nodes get deleted in the first round.

\section{Sketch Quality}\label{app:SketchQuality}

We now give few additional definitions that we use later in our proofs.
Let $\Lambda_{j}$ be the set  $\Lambda$ at the end of the $j$-th iteration of the algorithm and let $H_{\Lambda_j}$ be a graph that contains each edge of $G$ induced by $\Lambda_{j}$ with probability $p_{j+1}$. Lets define $H'_j=H_j \cup H_{\Lambda_{j-1}}$. Note that $H'_j$ contains each edge of $G$ with probability $p_j$. Furthermore for a graph $G$ we define $S_k$ as the subgraph of $G$ induced by the nodes with coreness number at least $k$ in $G$. We also denote $H'_{j,k}=H'_j\cap S_k$. Finally we define the graph $G_v$ as the subgraph of $G$ induced by the nodes removed after $v$ by the classic peeling algorithm when it is run on $V$ and we denote $H'_{j,v}=H'_j\cap G_v$. We are now ready to prove an upper bound and a lower bound on the coreness number of nodes that are added to $\Lambda$ in round $j$.

\begin{lemma}[Lemma~\ref{lm:levelj} restated]
For all $0\leq j\leq \log n$ such that $p_j \leq 1$ and for any node $v$ added to $\Lambda$ in round $j$ we have with probability $1-\frac{1}{3n}$ that: 
$C(v) < 2 (1+\epsilon)\frac{n}{2^{j-1}}$.

Furthermore for all $0\leq j\leq \log n$ such that $p_j < 1$ we have with probability $1-\frac{1}{3n}$ that: 
$C(v)\geq 2 (1-\epsilon)\frac{n}{2^{j}}  $.
\end{lemma}
\begin{proof}
By using the union bound and by fixing $\delta=\frac1n$ we have that Lemma~\ref{lm:cher0} and Lemma~\ref{lm:cher2} holds for all $H_{j,k}$ and $H_{j,v}$ for all choices of $j$ and $k$ and $v$ with probability $1-\frac{2 n \log n}{3 n^3} \geq 1-\frac{1}{3n}$. In the rest of the proof we assume that both lemma hold.
		
To prove the first statement of the theorem, we pick any vertex $v$ with $C(v)\geq 2 (1+\epsilon)\frac{n}{2^{j-1}}$ and show that $v$ is included in $\cup_{s=0}^{j-1}\Lambda_{s}$, and thus is not in $\Lambda_j$. Specifically, we show that if $v \notin \cup_{s=0}^{j-2}\Lambda_{s}$, then $v \in \Lambda_{j-1}$. Therefore, we have $v \notin \Lambda_j$, as desired.
		
Let $k=C(v)$. By applying Lemma \ref{lm:cher0} to $H'_{j-1,k}$ for any vertex $u \in H'_{j-1,k}$ we have
$|d_{H'_{j-1,k}}(u) - p_{j-1} d_{S_{k}}(u)|\leq \epsilon d_{H'_{j-1,k}}(u).$
This gives us $d_{H'_{j-1,k}}(u)  \geq \frac{p_{j-1} d_{S_{k}}(u)}{1+\epsilon} $. Furthermore note that each vertex in $S_{k}$ has degree at least $k \geq 2  (1+\epsilon)\frac{n}{2^{j-1}}$. Thus we have 
$ d_{H'_{j-1,k}}(u)   \geq \frac{p_{j-1} d_{S_{k}}(u)}{1+\epsilon} 
		\geq \frac{p_{j-1}  2  (1+\epsilon)\frac{n}{2^{j-1}}}{1+\epsilon}
		 = 2 p_{j-1}  \frac{n}{2^{j-1}} = \frac{192 \log n}{\epsilon^2}.$
		This means that the degree of all vertices in $H'_{j-1,k}$, including $v$ is at least $  \frac{192 \log n}{\epsilon^2}$. In addition we have $H'_{j-1,k} \subseteq H'_{j-1}$. Therefore, we have $C_{H'_{j-1}}(v) \geq   \frac{192 \log n}{\epsilon^2}$. Recall that $H'_{j-1}=H_{j-1} \cup H_{\Lambda_{j-2}}$. By applying Lemma \ref{lm:GoodGreed} we have $l_{j-1}(v) \geq C_{H'_{j-1}}(v) \geq   \frac{192 \log n}{\epsilon^2}$. Thus, we have $v \in \Lambda_{j-1}$, if  $v \notin \cup_{s=0}^{j-2}\Lambda_{s}$. This proves the that $C(v) < 2 (1+\epsilon)\frac{n}{2^{j-1}}$ for all $0\leq j\leq \log n$.
		

Now we show that the lower bound by contradiction. Without loss of generality suppose that the first vertex that contradict the lemma is in level $\Lambda_j$. Let $k$ be the minimum coreness number of any vertex in $\Lambda_j$ and let $v$ be the first vertex in $\Lambda_j$ removed by the peeling algorithm when it is run on the entire graph $G$. Note that we have $C(v)=k$. Now we assume by contradiction $k = C(v) < 2 (1-\epsilon) \frac{n}{2^{j}}$.
		
Let $v'$ be the first vertex in $\Lambda_j$ removed by Algorithm \ref{Alg:GreedyLambda} which received the same label as $v$. Let $\tilde{G}$ to be the subgraph of $G$ induced by the vertices in $\cup_{s=0}^{j-1} \Lambda_s$ and the vertices in $\Lambda_j$ that are removed after $v'$ including $v'$. 
Note that for any $u \in \tilde{G}$, $C(v) \leq C(u)$. So by definition of $v$, $\tilde{G} \subseteq G_v$.

Now by definition of $v$ and $G_v$, we have $d_{G_v}(v) \leq C(v) < 2 (1-\epsilon) \frac{n}{2^{j}}$. By applying Lemma~\ref{lm:cher0} to $H'_{j,v}$ either we have $d_{H'_{j,v}}(v) < 2 p_j (1-\epsilon)\frac{n}{2^{j}} $ or we have $|d_{H'_{j,v}}(v) - p_j d_{G_v}(v)|\leq \epsilon d_{H'_{j,v}}(v)$. The latter gives us
$d_{H'_{j,v}}(v)   \leq   \frac {p_j d_{G_v}(v)}{1-\epsilon} <  2 p_j \frac{n}{2^{j}}.$
Hence, in both cases we have $d_{H'_{j,v}}(v) < 2 p_j \frac{n}{2^{j}}= \frac{192 \log n}{\epsilon^2}$. But now, note that the degree of $v$ when it receive its label from Algorithm \ref{Alg:GreedyLambda} is bounded by its degree in $H'_{j,v}$. Hence, the label assigned to $v$ is strictly less than $ \frac{192 \log n}{\epsilon^2}$, which contradicts the existence of $v$ and completes the proof.		 	 
\end{proof}

We are ready to state the approximation guarantees of our sketch in Lemma~\ref{lm:acc} 
\begin{lemma}[Lemma~\ref{lm:acc} restated]
	Algorithm \ref{Alg:sketch} computes a $1-2\epsilon$ approximate core labeling, with probability $1-\frac{2}{3n}$.
\end{lemma}
\begin{proof}
This proof is similar to the proof of Lemma \ref{lm:levelj}. However, here we use Lemma \ref{lm:levelj} and in each fixed $\Lambda_j$ we bound the core-label of each vertex $v\in \Lambda_j$.
Here we assume the statement of Lemma \ref{lm:levelj} holds and the statements of Lemma \ref{lm:cher0} and \ref{lm:cher2} holds for $H_{j,k}$ and $H_{j,v}$ for all choices of $j$ and $k$ and $v$. Indeed, by fixing $\delta=\frac13$ these hold with probability $1-\frac{2  n \log n}{3 n^3} \geq 1-\frac{2}{3n}$.


	Pick an arbitrary $0\leq j\leq \log n$ such that $p_j < 1$, and an arbitrary vertex $v\in \Lambda_j$ and let $k =C(v)$.
	Note that from Lemma \ref{lm:levelj} we have that $k=C(v)\geq  2 (1-\epsilon) \frac{n}{2^j}$, and thus for each vertex $u\in S_{k}$ we have
	\begin{align}\label{eq:goodDeg}
	 d_{S_k}(u) \geq k \geq 2 (1-\epsilon) \frac{n}{2^j}.
	\end{align}
	
	Now, by applying the second statement of Lemma \ref{lm:cher0} to $H'_{j,k}$ for any vertex $u \in H'_{j,k}$ we have
	$|d_{H'_{j,k}}(u) - p_{j} d_{S_{k}}(u)|\leq \epsilon d_{H'_{j,k}}(u)$.
	This together with inequality \ref{eq:goodDeg} gives us 
	$
	d_{H'_{j,k}}(u)  \geq \frac{p_{j} d_{S_{k}}(u)}{1+\epsilon} \geq \frac{p_{j} }{1+\epsilon} k. 	
	$
	This means that the degree of all of the vertices in $H'_{j,k}$, including $v$ is at least $ \frac{p_{j} }{1+\epsilon} k$. In addition we have $H_{j,k} \subseteq H'_{j}$. Therefore, we have $C_{H'_{j}}(v) \geq \frac{p_{j} }{1+\epsilon} k$. Recall that $H'_{j}=H_{j} \cup H_{\Lambda_{j-1}}$. By applying Lemma \ref{lm:GoodGreed} we have $l_{j}(v) \geq C_{H'_{j}}(v) \geq   \frac{p_{j} }{1+\epsilon} k$. Thus, the label Algorithm \ref{Alg:sketch} assigns to $v$ is either at least $(1-\epsilon) \frac{l_{j}(v)}{p_j}  \geq   \frac{1-\epsilon }{1+\epsilon} k \geq (1-2\epsilon) k $ or $(1-\epsilon) \frac{n}{2^{j-1}}$. In the former case clearly the label that Algorithm \ref{Alg:sketch} assigns to $v$ is lower bounded by $(1-2\epsilon)C(v)$. Using Lemma \ref{lm:levelj}, we have $C(v)\leq 2 (1+\epsilon)\frac{n}{2^{j-1}}$, and thus, in latter case the label that Algorithm \ref{Alg:sketch} assigns to $v$ is lower bounded by $\frac{1-\epsilon }{1+\epsilon}C(v) \geq (1-2\epsilon)C(v)$. 
	
Now pick the first $0\leq j\leq \log n$ such that $p_j = 1$, and an arbitrary vertex $v\in \Lambda_j$ and let $k =C(v)$. First note that at the end of round $j$ we have $\Lambda = V$. Furthermore we have by Lemma~\ref{lm:GoodGreed} that $l_j(v)\geq C(v)$. Thus, also in this case the label Algorithm \ref{Alg:sketch} assign to $v$ is either at least $(1-\epsilon) \frac{l_{j}(v)}{p_j}  \geq   (1-\epsilon)  C(v) $ or $(1-\epsilon) \frac{n}{2^{j-1}}$. In the former case clearly the label that Algorithm \ref{Alg:sketch} assigns to $v$ is lower bounded by $(1-2\epsilon)C(v)$. Using Lemma \ref{lm:levelj}, we have $C(v)\leq 2 (1+\epsilon)\frac{n}{2^{j-1}}$, and thus, again in latter case the label that Algorithm \ref{Alg:sketch} assigns to $v$ is lower bounded by $\frac{1-\epsilon }{1+\epsilon}C(v) \geq (1-2\epsilon)C(v)$. This concludes the proof of the lower bound for the labels assigned by Algorithm~\ref{Alg:sketch}.


	Next we show that for an arbitrary round $0\leq j\leq \log n$ the label that Algorithm \ref{Alg:sketch} assigns to $v$ in round $j$ is upper bounded by $C(v)$.
	By the way of contradiction, lets assume there exist some $v$ such that the label that Algorithm \ref{Alg:sketch} assigns to $v$ is strictly more than $C(v)$. Without loss of generality, lets assume $v$ is the first  vertex removed by the peeling algorithm when it is run on $G$ and such that the label that Algorithm \ref{Alg:sketch} assigns to $v$ is strictly more than $C(v)$. Lets assume $v \in \Lambda_j$ and let $k = C(v)$.
		
	First note that $C(v)< \frac{2(1-\epsilon)n}{2^{j-1}}$, otherwise the label assigned to $v$ cannot be bigger than $C(v)$ by the condition in lines 18-20(Note that $\frac{2(1-\epsilon)n}{2^{j-1}} = (1-\epsilon)\frac{384\log n}{\epsilon^2 p_j}$). So in the rest of the proof we can restrict our attention to the case when $C(v)< \frac{2(1-\epsilon)n}{2^{j-1}}$. Furthermore note that by Lemma~\ref{lm:levelj} for any node $u$  in $\cup_{s=0}^{j-1} \Lambda_s$ have at least $C(u)\geq \frac{2(1-\epsilon)n}{2^{j-1}}$, so $C(v)\leq C(u)$, for all $u \in \cup_{s=0}^{j-1} \Lambda_s$. Now let $v'$ be the vertex in $\Lambda_j$ that is the first vertex removed by Algorithm \ref{Alg:GreedyLambda} and that received the same label as $v$ by this algorithm. Let $\tilde{G}$ to be the subgraph of $G$ induced by the vertices in $\cup_{s=0}^{j-1} \Lambda_s$ and the vertices in $\Lambda_j$ that are removed after $v'$ by Algorithm \ref{Alg:GreedyLambda}. 
	By the way of picking $v$, for any $u \in \tilde{G}$ we have $C(v) \leq C(u)$. Therefore, we have 
$\tilde{G} \subseteq G_{v}$.

	Now if $p_j = 1$, by Lemma~\ref{lm:GoodGreed} we get that the the label assigned to $v$ is equal to $C(v)$ and so we get a contradiction. So in the rest of the proof we assume that $p_j < 1$.

	In this case subgraph $H_j$ at the time of removing $v'$ is a subgraph of $H'_{j,v}$. In turn this implies  that $l_j(v) \leq d_{H'_{j,v}}(v)$. On the other hand, since $v \in \Lambda_j$ we have $l_j(v) \geq \frac{192 \log n}{\epsilon^2}$. Thus, we have $d_{H'_{j,v}}(v) \geq \frac{192 \log n}{\epsilon^2} $. Therefore, applying Lemma \ref{lm:cher2} to $H_{j,v}$ gives us
	$
	|d_{H'_{j,v}}(v) - p_j d_{G_v}(v)|\leq \epsilon d_{H'_{j,v}}(v).
	$
	By rearranging this we have 
	$
	d_{H'_{j,v}}(v) \leq \frac{p_j d_{G_v}(v)}{1-\epsilon}  = \frac{p_j k}{1-\epsilon} .
	$ 
	This together with $l_j(v) \leq d_{H'_{j,v}}(v)$ gives us $l_j(v) \leq \frac{p_j k}{1-\epsilon}$. 
	Now note that the label that Algorithm \ref{Alg:sketch} assigns to $v$ is upper bounded $\frac{(1-\epsilon) l_j(v)}{p_j}$. So we have that
	$
	\frac{(1-\epsilon) l_j(v)}{p_j} \leq \frac{(1-\epsilon) \frac{p_j k}{1-\epsilon}}{p_j} = k = C(v). 
	$
	So we know that for an arbitrary round $0\leq j\leq \log n$ such that $p_j < 1$ the label that Algorithm \ref{Alg:sketch} assigns to $v$ in round $j$ is between $(1-2\epsilon)C(v)$ and $C(v)$. This concludes the proof.
\end{proof}

\comment{
\section{A streaming algorithm in the turnstile settings}\label{app:turnstile}
Here we show an application of our sketch in the turnstile setting, where we have both addition and deletion of the edges. 

To handle deletions, we cannot use only the simple data structure presented in the previous section. We need to use in addition a $t$-sparse recovery data structure using $\tilde{O}(t)$ space~\cite{barkay2015efficient}. Before describing our algorithm we recall the definition of $t$-sparse recovery data structure.

\begin{definition}
	 A $t$-sparse recovery data structure is a data structure which handles insertions and deletions of elements in a set and such that if the current number of elements stored in it is at most $t$, then these elements can be recovered with high probability.
\end{definition}

Now we can describe our algorithm,  for each graph $H_j$ and each $v$ vertex in $H_j$ we keep a $\frac{24\log n}{\epsilon^2}$-sparse recovery data structure $S^v_j$. Upon insertion of each edge $(u,v)$ if it is sampled in the $j$-th iteration (passed Line~\ref{line:probIf} of Algorithm~\ref{Alg:turnstile-in}) and is not inserted in $H_j$ (i.e., failed Line~\ref{line:lambdaIf} of Algorithm~\ref{Alg:turnstile-in}), we insert $(u,v)$ to $S^u_j$ and $S^v_j$.

During the duration of the algorithm we need to remember the random number $r$ that is associated to the edge $(u,v)$. Unfortunately we can't store this number explicitly, although  we can use a single $\tilde{O}(n)$-minwise independent hash function~\cite{feigenblat2011exponential} to hash all edges to their corresponding random numbers. The pseudo-code our algorithm in the turnstile setting is presented in Algorithm~\ref{Alg:turnstile-in} and Algorithm~\ref{Alg:turnstile-del}.

Edge insertion is similar to the insertion only case with the only two modification described above.

Handling edge deletion is slightly more complicated, this is true because they could potentially cause chains of updates in the data structures. In the deletion step, if $(u,v)$ was not sampled in the $j$-th iteration, we do not need to do anything. Otherwise, if $(u,v)\notin H_j$, it means that it is inserted into $S^u_j$ and $S^v_j$. In this case we delete $(u,v)$ from $S^u_j$ and $S^v_j$ and we are done.

Now if $(u,v)\in H_j$, we need to recompute $\Lambda_{j}$, and denote it by $\Lambda_{j}^{new}$.

Let $\Delta_j = \Lambda_{j} - \Lambda_{j}^{new}$ be the sequence of vertices that are removed from $\Lambda_{j}$ after this recomputation, ordered by their removal order in the greedy algorithm (See Line \ref{line:delta}). We iterate over the vertices in $\Delta_j$ and recover their edges them (See Line \ref{line:recover}). 
Remember that the edges in $S_{j+1}^{v'}$ are those that both their endpoints fall in $\cup_{i\leq j}\Lambda_i$. Let $\Delta_j=\{v'_1,v'_2,v'_3,\dots,v'_{k}\}$ be the sequence vertices that move out of $\Lambda_j$ as we remove $(u,v)$, and assume $v'_1,v'_2,v'_3,\dots,v'_{k}$ is in the same order that the greedy algorithm (in Line \ref{line:greedy}) removes these vertices. Recall that, if a vertex is not in $\Lambda_j^{new}$ its label assigned by the greedy algorithm is at most $\frac{24\log n}{\epsilon^2}$ (due to Line \ref{line:lj}). Thus, upon removal of $v'_1$ the number of edges between $v'_1$ and (the old) $\cup_{i\leq j}\Lambda_i$ is at most $\frac{24\log n}{\epsilon^2}$. Thus, we can recover all of the edges of $v'_1$ using our sparse recovery data structure $S_{j+1}^{v'}$ and add it to $H_{j+1}$. Similarly note that, when greedy is removing $v'_2$ the number of edges between $v'_2$ and $Lambda_j - \{v_1 \}$  is at most $\frac{24\log n}{\epsilon^2}$. Note that after recovering $S_j^{v'_1}$, if there is an edge between $v'_1$ and $v'_2$ or not, we removed it from $S_j^{v'_2}$. Thus, $S_j^{v'_2}$ only contains the edges between $v'_2$ and $\Lambda_j - \{v_1 \}$ which is upper bounded by $\frac{24\log n}{\epsilon^2}$. Hence we can recover the edges in $S_j^{v'_2}$. The same argument holds for $v_3$ and so on.

By the results presented in the previous section
we obtain the following corollary.

\begin{corollary}[Corollary~\ref{cor:turnstile} restated]
	There exists a turnstile algorithm that computes an approximate core-labeling of the input graph using $\tilde{O}(n)$ space.
\end{corollary}

\begin{algorithm}[ht!]
\begin{algorithmic}[1]
	\small
	\STATE \textbf{Input:} A sequence of insertions and deletions of the edges of the input graph $G(V,E)$ with $n$ vertices and parameter $\epsilon \in (0,1]$.
	\STATE Initialize $\Lambda_{i} \leftarrow \emptyset$, $\forall i$
	\STATE $p_i \leftarrow 2^i \frac{12 \log n}{\epsilon^2 n}$, $\forall i$
	\FOR{$j = 0$ to $\log n$}
		\FOR{$v \in V$}
			\STATE Initiate a $\frac{24\log n}{\epsilon^2}$-sparse recovery $S^v_j \leftarrow \emptyset$
		\ENDFOR
	\ENDFOR
	 \STATE $\Lambda_{-1} \leftarrow \emptyset$
	 \STATE Let $h(u,v): (V,V)\rightarrow [0,1]$ be a $\tilde{O}(n)$-minwise independent hash function.

	\STATE  \textbf{Insertion of $(u,v)$}
	\STATE $r \leftarrow h(u,v)$
	\FOR{$j = 0$ to $\log n$}
		\IF{$r\leq p_j$}\label{line:probIf}
			\IF{$v\notin\cup_{i<j}\Lambda_{i}$ or $u \notin \cup_{i<j}\Lambda_{i}$}\label{line:lambdaIf}
				\STATE Add $(u,v)$ to $H_j$
				\STATE Run $Exclusive\_Core\_Labeling(H_j,\cup_{i<j}\Lambda_{i})$ and denote the label of vertex $i$ on $H_j$ by $l_j(i)$
				\STATE $\Lambda_{j}\leftarrow \emptyset$
				\FOR{$i \in H_j$}
					\IF{$l_j(i) \geq \frac{24\log n}{\epsilon^2} \vee p_j = 1$}
						\IF{$l_j(i) \leq \frac{48\log n}{\epsilon^2}$}
							\STATE Set the label of vertex $i$ to $(1-\epsilon)\frac{l_j(i)}{p_j}$
							\STATE Add $i$ to $\Lambda_{j}$
						\ELSE
							\STATE Set the label of vertex $i$ to $ \frac{2(1-\epsilon)n}{2^{j-1}}$
							\STATE Add $i$ to $\Lambda_{j}$
						\ENDIF
					\ENDIF
				\ENDFOR
			\ELSE
				\STATE Add $(u,v)$ to $S^u_j$
				\STATE Add $(u,v)$ to $S^v_j$
			\ENDIF
		\ENDIF
	\ENDFOR
	\caption{\small Pseudocode to handle insertion to compute $1-O(\epsilon)$ approximate core-labeling in the turnstile setting.}
	\label{Alg:turnstile-in}
\end{algorithmic}
\end{algorithm}
\begin{algorithm}[ht!]
\begin{algorithmic}[1]
\STATE \textbf{Deletion of $(u,v)$}\;
\STATE $r \leftarrow h(u,v)$\; 
\FOR{$j = 0$ to $\log n$}
	\IF{$r\leq p_j$}
		\IF{$(u,v) \in H_j$}
			\STATE Delete $(u,v)$ from $H_j$\;
			\STATE Run $Exclusive\_Core\_Labeling(H_j,\Lambda_{j-1})$ and denote the label of vertex $i$ on $H_j$ by $l_j(i)$\;\label{line:greedy}
			\STATE $\Lambda_{j}^{new}\leftarrow \emptyset$\;
			\FOR{$i \in H_j$}
				\IF{$l_j(i) \geq \frac{24\log n}{\epsilon^2} \vee p_j = 1$}\label{line:lj}
					\IF{$l_j(i) \leq \frac{48\log n}{\epsilon^2}$}
						\STATE Set the label of vertex $i$ to $(1-\epsilon)\frac{l_j(i)}{p_j}$\;
						\STATE Add $i$ to $\Lambda_{j}^{new}$\;
					\ELSE
						\STATE Set the label of vertex $i$ to $ \frac{2(1-\epsilon)n}{2^{j-1}}$\;
						\STATE Add $i$ to $\Lambda_{j}^{new}$\;
					\ENDIF
				\ENDIF
				\ENDFOR
				\STATE Let $\Lambda_j^{\Delta} \leftarrow \Lambda_{j} - \Lambda_{j}^{new}$ ordered by the removal oreder in Line \ref{line:greedy}\;\label{line:delta}
				\FOR{$v'\in \Lambda_j^{\Delta}$}\label{line:recover}
					\STATE Recover $S^{v'}_j$\;
					\STATE Add the edges in $S^{v'}_j$ to $H_j$\;
					\STATE Remove all the edges in $S^{v'}_j$ from the sparse recovery of both their endpoints\;
				\ENDFOR
				\STATE $\Lambda_{j} \leftarrow \Lambda_{j}^{new}$\;
			
		\ELSE
			\STATE Delete $(u,v)$ from $S^u_j$
			\STATE Delete $(u,v)$ from $S^v_j$
		\ENDIF
	\ENDIF	
\ENDFOR

	\caption{\small Pseudocode to handle deletion to compute $1-O(\epsilon)$ approximate core-labeling in the turnstile setting.}
	\label{Alg:turnstile-del}
\end{algorithmic}
\end{algorithm}
}
\end{appendix}

\end{document}